\documentclass[final,5p,times,numbers]{elsarticle}

\usepackage{amssymb}
\usepackage{lipsum}
\usepackage{amsmath}
\usepackage{booktabs} 
\usepackage{multirow} 
\usepackage[version=4]{mhchem} 
\usepackage{makecell}

 \usepackage{booktabs}
 \usepackage{threeparttable}
 \usepackage{array}
 \usepackage{xurl}

 \usepackage{graphicx} 

\providecommand{\A}{\AA}

\begin{document}
\begin{frontmatter}
\title{Unraveling Lithium Storage and Defect-Modulated Transport in 2D Copper Boride: From First Principles to Machine-Learned Molecular Dynamics}

\author[first]{Subhasis Sarkar}
\author[first]{Rajnendra Singh}

\author[second]{Brahmananda Chakraborty\corref{cor1}}
\ead{brahma@barc.gov.in}

\author[first]{Sridhar Sahu\corref{cor1}}
\ead{sridharsahu@iitism.ac.in}

\affiliation[first]{
    organization={Computational Materials Research Lab,
        Department of Physics,
        Indian Institute of Technology (Indian School of Mines)},
    city={Dhanbad},
    postcode={826004},
    country={India}
}

\affiliation[second]{
    organization={High Pressure and Synchrotron Radiation Physics Division,
        Bhabha Atomic Research Centre},
    city={Mumbai},
    postcode={400085},
    country={India}
}

\cortext[cor1]{Corresponding authors}
            
\begin{abstract}
In this work, we combine first-principles calculations, ab initio
molecular dynamics (AIMD), and machine-learning-interatomic-potential
molecular dynamics (MLIP-MD) to investigate lithium storage and
transport in two-dimensional Cu$_8$B$_{14}$, including an
experimentally identified line-defect configuration. Pristine
Cu$_8$B$_{14}$ remains metallic upon lithiation and exhibits
favourable Li adsorption, a single-surface theoretical capacity of
427~mAh~g$^{-1}$, and an average open-circuit voltage of
approximately 0.53~V. The lowest Li migration barrier on the pristine
surface is 0.32~eV. The line-defected structure retains a capacity of
approximately 400~mAh~g$^{-1}$ while reducing the local migration
barrier to 0.21~eV. To access Li dynamics beyond the picosecond time
scale of AIMD, a pretrained MACE-MP-0 model was fine-tuned against
system-specific DFT energies and forces and employed in
nanosecond-scale simulations from 400 to 600~K. Both pristine and
line-defected lithiated monolayers remain structurally stable over
this temperature range. At 400~K, the line-defected system exhibits
a higher in-plane Li tracer diffusivity
($3.97\times10^{-5}$~cm$^{2}$~s$^{-1}$) than the pristine
monolayer ($2.50\times10^{-5}$~cm$^{2}$~s$^{-1}$), whereas the
pristine system becomes more diffusive at 500 and 600~K. Arrhenius
analysis yields effective activation energies of 0.147 and 0.225~eV
for the line-defected and pristine systems, respectively. These
results show that the line defect modifies the Li-transport landscape
and its temperature dependence rather than uniformly enhancing
long-range diffusion, while preserving competitive Li-storage
capacity and structural stability.

\end{abstract}

\begin{keyword}

Copper boride \sep lithium-ion batteries \sep anode materials \sep density functional theory (DFT) \sep machine-learning-interatomic-
potential (MLIP)   \sep line defect \sep \textit{ab initio} molecular dynamics (AIMD)

\end{keyword}

\end{frontmatter}

\section{Introduction}
\label{introduction}

The intermittency of renewable energy sources, such as wind or solar, necessitates robust energy storage solutions to ensure the grid reliability \cite{i1,i2}. Among the diverse storage technologies, lithium-ion batteries (LIBs) have established dominance in portable electronics \cite{i3}, electric vehicles (EVs) \cite{i4}, and stationary grid storage \cite{i5}. This widespread adoption is attributed to the high energy density, favourable operating voltage, and long-term cycling stability of LIBs \cite{i6,i7}.
However, the performance of state-of-the-art LIBs is approaching its physicochemical limits. The commercial standard anode, graphite, is restricted by a theoretical specific capacity of 372 mAh/g and poor rate capability under high-current charging due to sluggish intercalation kinetics\cite{i33}. Consequently, there is an urgent demand for next-generation anode materials that possess superior specific capacity, enhanced electrical conductivity, and rapid ion diffusion kinetics \cite{i10,i11}.

The advent of graphene catalyzed the intensive exploration of two-dimensional materials for electrochemical applications. Following graphene, a wide array of 2D materials including transition metal oxides (TMOs), transition metal dichalcogenides (TMDs), MXenes, and monoelemental Xenes have been investigated as high-performance battery electrodes \cite{i12,i13,i14,i15,i16,i17}. While many of these materials offer capacities exceeding that of graphite, they often face critical bottlenecks, such as the low intrinsic electrical conductivity common in TMOs and TMDs, or issues related to structural instability and sheet aggregation which severely limit their cycle life and practical application\cite{i34}.

Recently, 2D metal borides (MBenes) have emerged as a formidable class of materials 
capable of addressing these limitations \cite{i18,i19}. Unlike many semiconducting 2D materials, metal borides typically exhibit metallic conductivity, 
high mechanical hardness, and excellent chemical stability \cite{i35} attributes 
essential for electrode longevity and rate performance.

Current literature on MBenes, such as 2d Mo$_2$B$_{2}$ and Fe$_2$B$_{2}$, suggest that they have promising result regarding low diffusion energy barriers and high Li storage capacities \cite{T4}. MBenes also hold promise in catalytic applications, including the hydrogen evolution reaction (HER) and nitrogen reduction reaction (NRR), highlighting their multifunctional electrochemical activity \cite{i25,i27,T4,i29,i28,i37}.

A promising development is the recent synthesis of the $\text{Cu}_8\text{B}_{14}$ monolayer, which has been identified as a dynamically stable structure at the atomic scale \cite{i22,i31,i32}. This material presents a unique characteristic, combining the lightweight and high-capacity potential of boron with the high conductivity of copper. Despite this potential, a systematic investigation into the electrochemical performance of $\text{Cu}_8\text{B}_{14}$ as an anode for Li-ion batteries is currently lacking.

In this study, we employ first-principles density functional theory (DFT) calculations to systematically evaluate the suitability of the $\text{Cu}_8\text{B}_{14}$ monolayer as an anode material for next-generation Li-ion batteries. We rigorously examine its structural and thermal stability using ab initio molecular dynamics (AIMD) simulations. Furthermore, we simulated the interaction mechanism of lithium adatoms with the $\text{Cu}_8\text{B}_{14}$ surface by calculating adsorption energies, theoretical storage capacities, and open-circuit voltage (OCV) profiles. Finally, ionic mobility is assessed by mapping the diffusion energy barriers along various migration pathways using the nudged elastic band (NEB) method. Our findings indicate that $\text{Cu}_8\text{B}_{14}$ exhibits a good balance of high specific capacity and low diffusion barriers, positioning it as a highly promising candidate for advanced energy storage applications.

\section{Computational Methodology}

\subsection{First-principles DFT calculations}

First-principles calculations were performed within density
functional theory (DFT) using the projector-augmented-wave (PAW)
method as implemented in the Vienna \textit{Ab initio} Simulation
Package (VASP)~\cite{m3}. Exchange-correlation effects were treated within the
generalized-gradient approximation (GGA) using the
Perdew-Burke-Ernzerhof (PBE) functional~\cite{m4}., while dispersion interactions were accounted for
using Grimme's DFT-D3 correction~\cite{m5,m6}. A plane-wave
kinetic-energy cutoff of $500~\mathrm{eV}$ was employed for the
standard structural and electronic calculations. The electronic
self-consistency criterion was set to $10^{-5}~\mathrm{eV}$,
and structural relaxations were continued until the residual atomic
forces were below $0.02~\mathrm{eV\,\AA^{-1}}$. Spin-polarized
calculations were initially performed to examine the possible
presence of magnetic ground states. Since the pristine and
lithiated structures converged to non-magnetic states, subsequent
calculations were performed without spin polarization.

To simulate the 2D material, a vacuum spacing of $30~\mathrm{\AA}$ was introduced along the
$z$ direction to minimize interactions between periodically repeated
monolayers. Brillouin-zone integrations were performed using
Monkhorst--Pack $k$-point meshes \cite{m7}, with the sampling density
adjusted according to the simulation cell. For the rectangular
Cu$_8$B$_{14}$ unit cell, $7\times3\times1$ and $9\times5\times1$
meshes were employed for structural optimization and static electronic
calculations, respectively. For the oblique $4\times1$ supercell,
the corresponding meshes were $3\times3\times1$ and
$5\times5\times1$, respectively.

To evaluate the dynamical stability of the Cu$_8$B$_{14}$
monolayer, phonon dispersion relations were calculated using the
finite-displacement method as implemented in Phonopy
\cite{M13}, with force constants constructed using the
symmetry-adapted force-constant (symfc) backend \cite{M14}.
A $3\times1\times1$ supercell was constructed from the fully
relaxed rectangular unit cell. Because phonon frequencies are
particularly sensitive to numerical noise in the atomic forces,
more stringent convergence parameters were employed than in the
standard structural calculations. The plane-wave cutoff energy was
increased to $600~\mathrm{eV}$ and the electronic self-consistency
criterion was tightened to $10^{-8}~\mathrm{eV}$. Prior to the
finite-displacement calculations, structural relaxation was
performed using a force-convergence criterion of
$10^{-4}~\mathrm{eV\,\AA^{-1}}$. The projection operators were
evaluated in reciprocal space to further improve the numerical
accuracy of the calculated forces.

Lithium migration barriers were calculated using the nudged elastic
band (NEB) method as implemented through the VTST toolkit
\cite{m8,m9,web1}. Five intermediate images were generated between
the initial and final local minima, with a spring constant of
$5~\mathrm{eV\,\AA^{-2}}$.

Ab initio molecular dynamics (AIMD) simulations were performed
with a time step of $1~\mathrm{fs}$ to assess finite-temperature
stability and short-time Li dynamics. For the pristine
Cu$_8$B$_{14}$ monolayer, the system was heated from 0 to
$400~\mathrm{K}$ over $3~\mathrm{ps}$ using velocity scaling,
followed by a $5~\mathrm{ps}$ trajectory at $400~\mathrm{K}$
in the canonical (NVT) ensemble using a Nos\'e--Hoover
thermostat \cite{m11}. Additional AIMD simulations of the
lithiated pristine and line-defected systems were performed at
$400~\mathrm{K}$ in the NVT ensemble to examine their thermal
stability and short-time Li dynamics. The resulting Li-containing
trajectories were analysed using the common displacement-analysis
procedure described in Section~\ref{sec:diffusion_analysis}.

\subsection{Machine-learning interatomic potential}

To overcome the limited time scales accessible to ab initio
molecular dynamics (AIMD) and to investigate long-time Li-ion
dynamics in the lithiated Cu--B monolayers, we employed
machine-learning interatomic-potential-driven molecular dynamics
(MLIP-MD). The interatomic potential was constructed within the
MACE (Multi-Atomic Cluster Expansion) framework, an equivariant
message-passing architecture capable of representing high-order
many-body interactions \cite{Batatia2022MACE,Kovacs2023MACE}.
Rather than training potential from scratch, the pretrained
MACE-MP-0 medium foundation model was fine-tuned using
system-specific DFT energies and atomic forces for Cu--B--Li
configurations \cite{Batatia2025Foundation}. This transfer-learning strategy leverages the representations learned
by the foundation model while adapting the potential to the local
environments relevant to Li adsorption and diffusion in the pristine
and line-defected Cu--B monolayers. Similar
fine-tuning of MACE-MP foundation models has recently been
employed to access nanosecond-scale Li diffusion in electrode
materials \cite{m15}.

The system-specific reference dataset was constructed from
finite-temperature DFT trajectories representing four structural
classes: pristine Cu--B, lithiated pristine Cu--B containing
42 Li atoms, line-defected Cu--B, and the corresponding lithiated
line-defect structure containing 26 Li atoms. Following exclusion
of the initial equilibration portions of the trajectories, candidate
configurations were generated to provide broad coverage of the
sampled configurational space, including structures representative
of different trajectory intervals as well as energy- and
force-stratified configurations. To reduce the redundancy arising
from strongly correlated molecular-dynamics frames, the candidate
structures were represented using the Smooth Overlap of Atomic
Positions (SOAP) descriptor \cite{Bartok2013SOAP} and
down-selected by farthest-point sampling (FPS)
\cite{Imbalzano2018FPS}, thereby preferentially retaining
configurationally diverse structures. For the Li-containing
systems, configurations associated with identified Li-motion
events were additionally retained to ensure that local environments
relevant to Li migration were represented in the reference data.
The resulting dataset comprised 230 pristine, 270
pristine+42Li, 230 line-defected, and 270 line-defect+26Li
configurations, giving a total of 1000 DFT-labelled structures.
The DFT total energies and atomic forces of these configurations
were used as reference quantities for fine-tuning. The dataset
was subsequently divided into training, validation, and test
subsets in an 80:10:10 ratio while preserving the relative
representation of the four structural classes, resulting in 800,
100, and 100 configurations, respectively.

The pretrained MACE-MP-0 medium model was fine-tuned on the
system-specific Cu--B--Li reference dataset, with all trainable
model parameters allowed to adapt to the DFT reference data. The
medium MACE architecture employs two equivariant message-passing
layers with correlation order 3, spherical-harmonic features up to
$l_{\mathrm{max}}=3$, and a radial cutoff of $6.0~\mathrm{\AA}$,
corresponding to an effective receptive field of approximately
$12~\mathrm{\AA}$ through the two message-passing steps
\cite{Batatia2025Foundation}. Fine-tuning was performed for
100 epochs using the Adam optimizer \cite{Kingma2015Adam} with
an initial learning rate of $5\times10^{-4}$. Energies and atomic
forces were optimized simultaneously using relative loss weights
of 1 and 10, respectively, placing greater emphasis on the force
information required for molecular-dynamics simulations.

The predictive performance of the potential was assessed at two
levels. First, both the pretrained MACE-MP-0 medium model and the
fine-tuned MACE-MP-0 model were evaluated on the same held-out test set
containing 100 configurations. As summarized in
Table~\ref{tab:mace_validation}, system-specific fine-tuning reduced
the energy RMSE from $187.8$ to $11.2~\mathrm{meV\,atom^{-1}}$
and the force RMSE from $273.0$ to
$58.2~\mathrm{meV\,\AA^{-1}}$. To further assess transferability
beyond the configurations used during model development, the
fine-tuned MACE-MP-0 model was evaluated against an independent
6000-frame DFT-AIMD trajectory of the line-defected structure
containing 27 Li atoms. For this independent trajectory, the
energy and all-species force RMSEs were
$4.46~\mathrm{meV\,atom^{-1}}$ and
$83.27~\mathrm{meV\,\AA^{-1}}$, respectively. The Li-specific
force RMSE was $46.22~\mathrm{meV\,\AA^{-1}}$, while the mean
cosine similarity between the DFT and MACE Li-force vectors was
0.97, indicating that the force directions governing Li motion
were well reproduced in configurations outside the model-development
dataset.

\begin{table*}[t]
\centering
\caption{Validation of the pretrained MACE-MP-0 medium model and the
system-specific fine-tuned MACE-MP-0 model against DFT reference data.
Energy and force errors are reported as RMSE/MAE. The independent
defect+27Li AIMD trajectory was not included in the training,
validation, or test subsets. A cosine similarity of unity corresponds
to perfect directional agreement between the DFT and predicted
Li-force vectors.}
\label{tab:mace_validation}

\footnotesize
\setlength{\tabcolsep}{4pt}

\begin{tabular*}{\textwidth}
{@{\extracolsep{\fill}}llcccc}
\hline
Evaluation set &
Model &
\shortstack{Energy error\\RMSE/MAE\\(meV atom$^{-1}$)} &
\shortstack{Force error\\RMSE/MAE\\(meV \AA$^{-1}$)} &
\shortstack{Li-force error\\RMSE/MAE\\(meV \AA$^{-1}$)} &
\shortstack{Mean Li-force\\cosine similarity} \\
\hline

Held-out test set &
Pretrained MACE-MP-0 &
187.78/165.93 &
273.03/185.66 &
-- &
-- \\

Held-out test set &
Fine-tuned MACE-MP-0 &
11.18/8.67 &
58.22/42.28 &
-- &
-- \\

Independent defect+27Li AIMD &
Fine-tuned MACE-MP-0 &
4.46/4.30 &
83.27/60.75 &
46.22/35.71 &
0.97 \\

\hline
\end{tabular*}
\end{table*}

\subsection{Molecular dynamics and diffusion analysis}
\label{sec:diffusion_analysis}

Long-time molecular dynamics simulations were performed using the
fine-tuned MACE-MP-0 potential to investigate the temperature dependence
of Li diffusion in lithiated pristine and line-defected Cu--B monolayers.
The pristine and line-defected systems contained 42 and 27 Li atoms,
respectively. Simulations were conducted at 400, 500, and
$600~\mathrm{K}$ in the canonical (NVT) ensemble using a
Nos\'e--Hoover-chain thermostat \cite{m11}, as implemented in the
Atomic Simulation Environment (ASE) \cite{Larsen2017ASE}.
For each system at each temperature, the trajectory was propagated
for $1~\mathrm{ns}$ with a time step of $1~\mathrm{fs}$.

Li self-diffusion was quantified from the time-origin-averaged in-plane
mean-squared displacement (MSD) using the Einstein relation
\cite{Purtscher2024Graphite,d3,m15},

\begin{equation}
D_{\mathrm{Li}}^{xy}
=
\frac{1}{2d}
\lim_{t\rightarrow\infty}
\frac{d}{dt}
\left[
\frac{1}{N_{\mathrm{Li}}}
\sum_{i=1}^{N_{\mathrm{Li}}}
\left\langle
\left|
\mathbf{r}^{xy}_{i}(t_{0}+t)
-
\mathbf{r}^{xy}_{i}(t_{0})
\right|^{2}
\right\rangle_{t_{0}}
\right],
\label{eq:einstein_diffusion}
\end{equation}

where $N_{\mathrm{Li}}$ is the number of Li atoms, $t_{0}$ denotes the
time origin, $t$ is the lag time, and $d=2$ is the dimensionality
of diffusion in the present two-dimensional systems. The quantity in
square brackets therefore corresponds to the time-origin-averaged
in-plane MSD,
$\mathrm{MSD}_{xy}(t)=\mathrm{MSD}_{x}(t)+
\mathrm{MSD}_{y}(t)$.
Atomic trajectories were first unwrapped across periodic boundaries,
and the mass-weighted center-of-mass displacement of the Cu--B
framework was removed before evaluating the Li displacements.
Diffusion coefficients were obtained from the slope of the linear,
near-Fickian region of the $\mathrm{MSD}_{xy}(t)$ curves.

The temperature dependence of the Li diffusivity was analyzed using
the Arrhenius relation,

\begin{equation}
D_{\mathrm{Li}}^{xy}(T)
=
D_{0}
\exp\left(
-\frac{E_{\mathrm{a}}}{k_{\mathrm{B}}T}
\right),
\label{eq:arrhenius_diffusion}
\end{equation}

where $D_{0}$ is the pre-exponential factor, $E_{\mathrm{a}}$ is the
effective activation energy for Li diffusion, $k_{\mathrm{B}}$ is the
Boltzmann constant, and $T$ is the absolute temperature.

\section{Results and Discussion}

\subsection{Structural and electronic properties of pristine Cu$_{8}$B$_{14}$}

\begin{figure}
    \centering 
    \includegraphics[width=0.5
    \textwidth, angle=0]{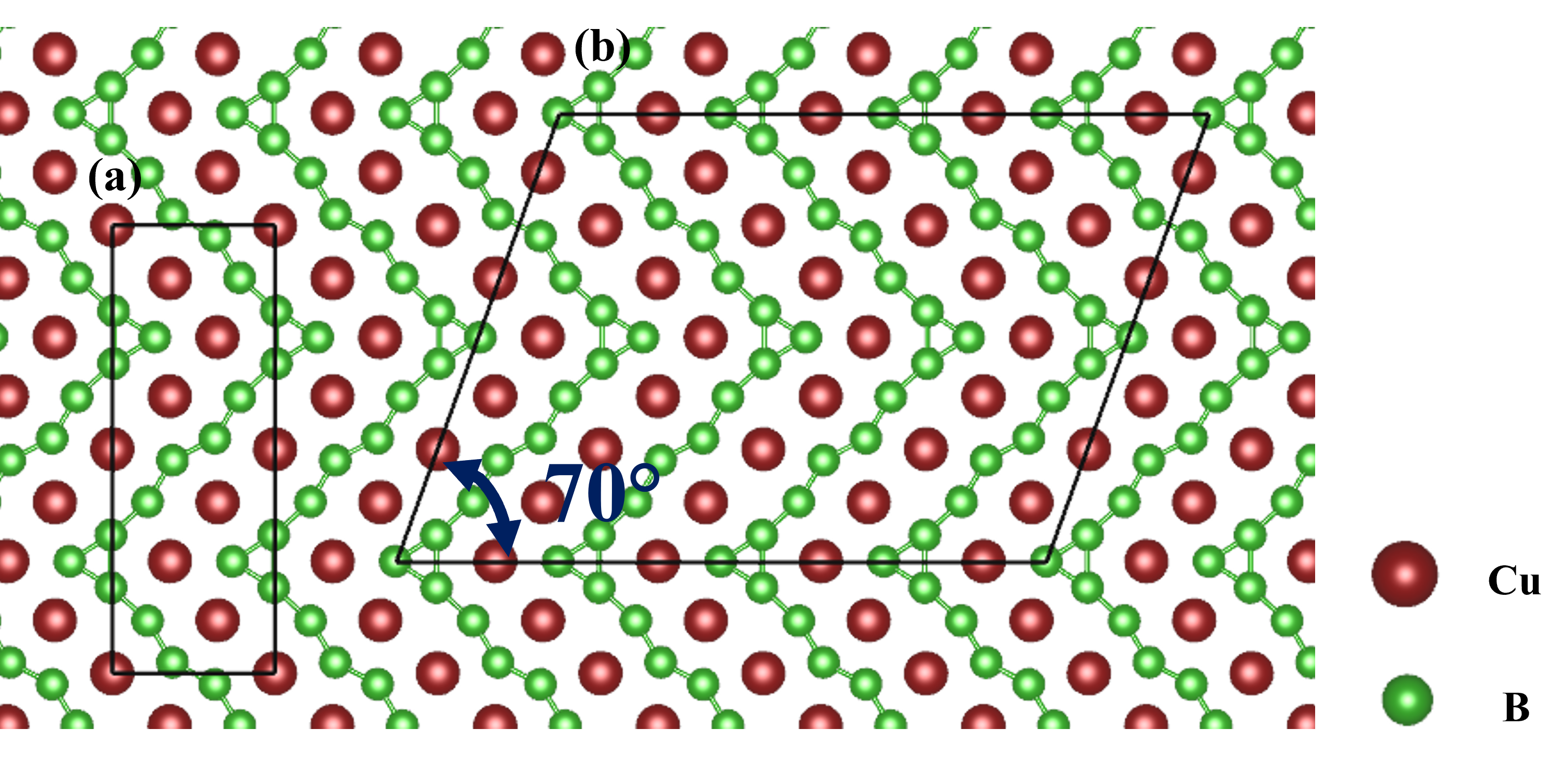}    
    \caption{(a) Top view of the rectangular unit cell and (b) The oblique $4 \times 1$ supercell of the Cu$_{8}$B$_{14}$ monolayer.} 
    \label{fig:structures}%
\end{figure}

\begin{figure}
    \centering 
    \includegraphics[width=0.53\textwidth, angle=0]{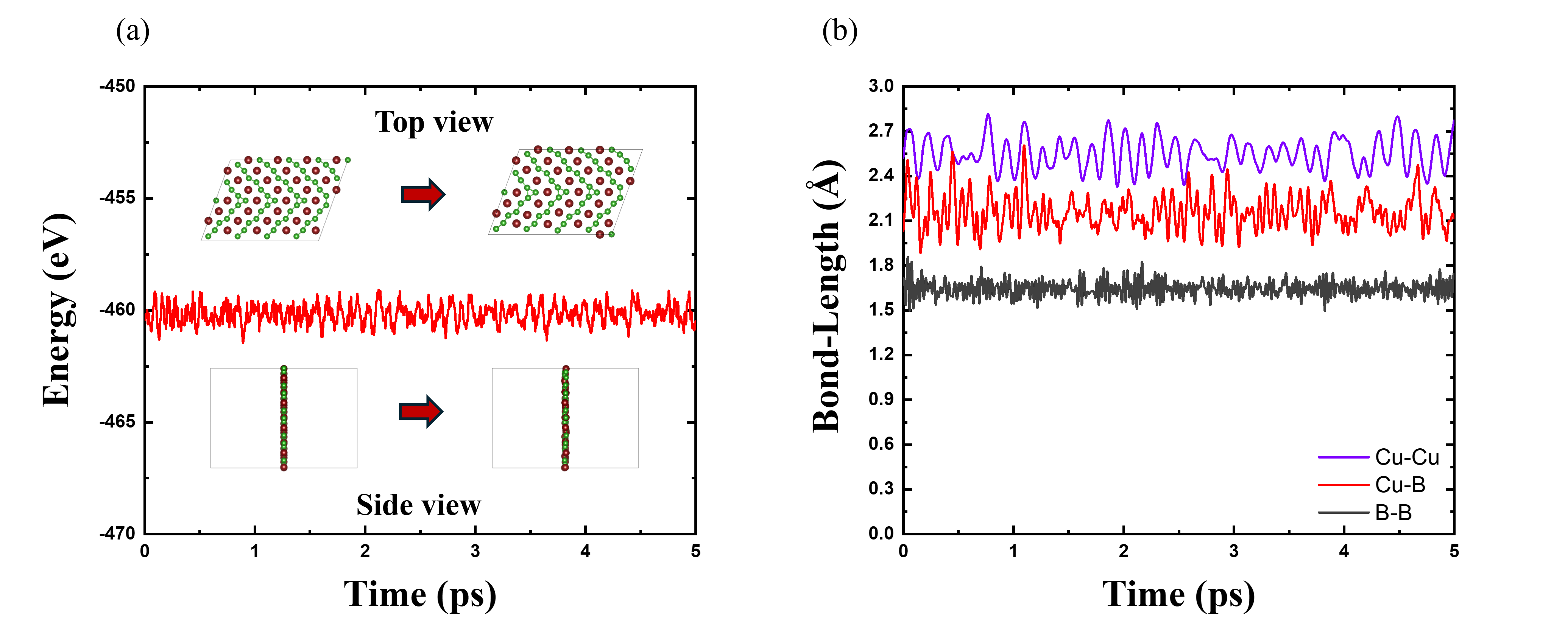}    
    \caption{(a) Evolution of total energy during the AIMD simulation at 400~K, with the inset showing the initial and final structures. (b) Bond length variation of Cu--Cu, Cu--B, and B--B with time throughout the simulation trajectory.} 
    \label{fig:aimd_pristine}
\end{figure}

\begin{figure}
    \centering 
    \includegraphics[width=0.42\textwidth, angle=0]{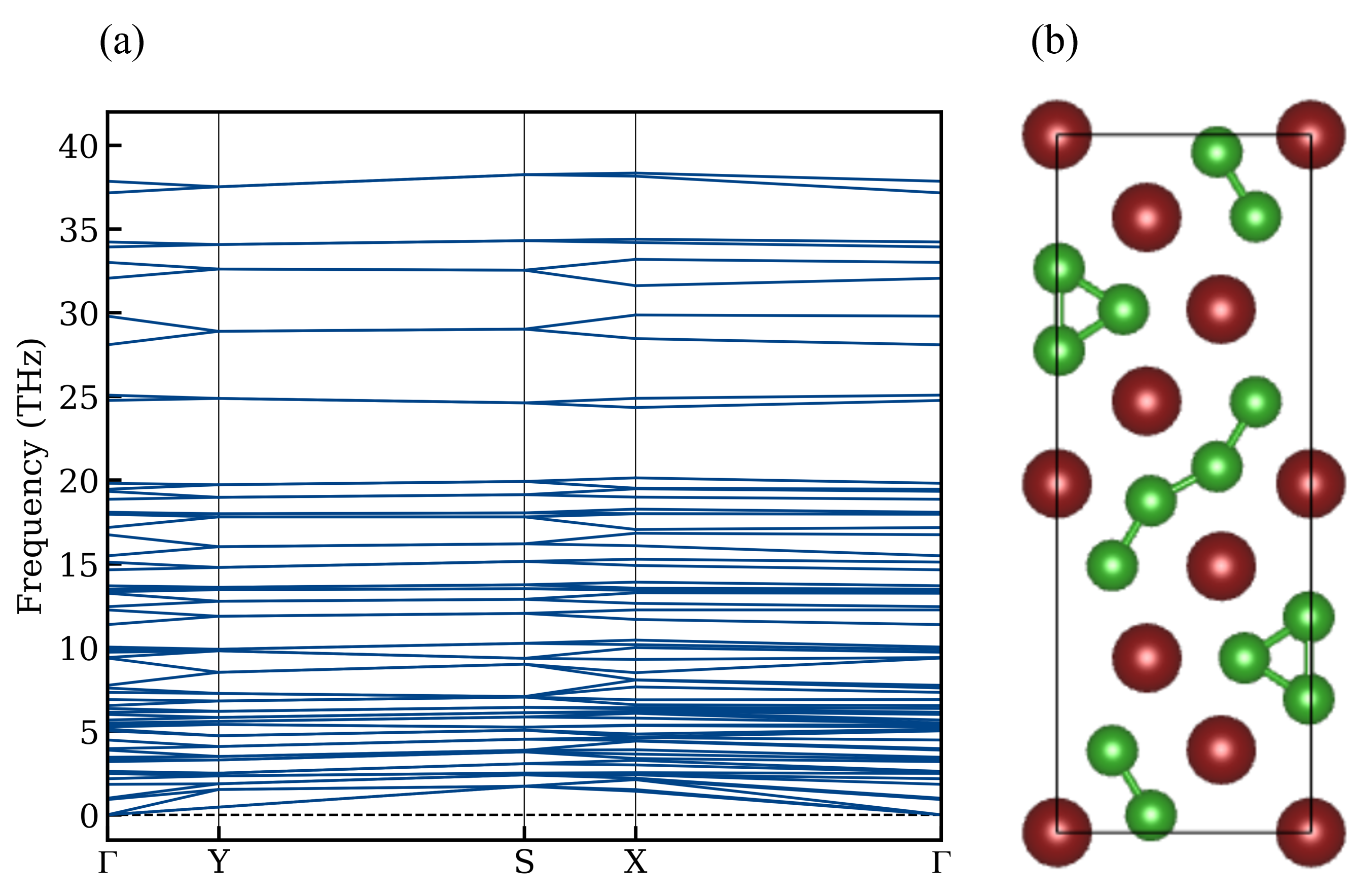}    
    \caption{(a) Phonon dispersion curve of the Cu$_{8}$B$_{14}$ monolayer (b) the structural model.} 
    \label{fig:phonon-dispersion}
\end{figure}

\begin{figure}
    \centering 
    \includegraphics[width=0.5
    \textwidth, angle=0]{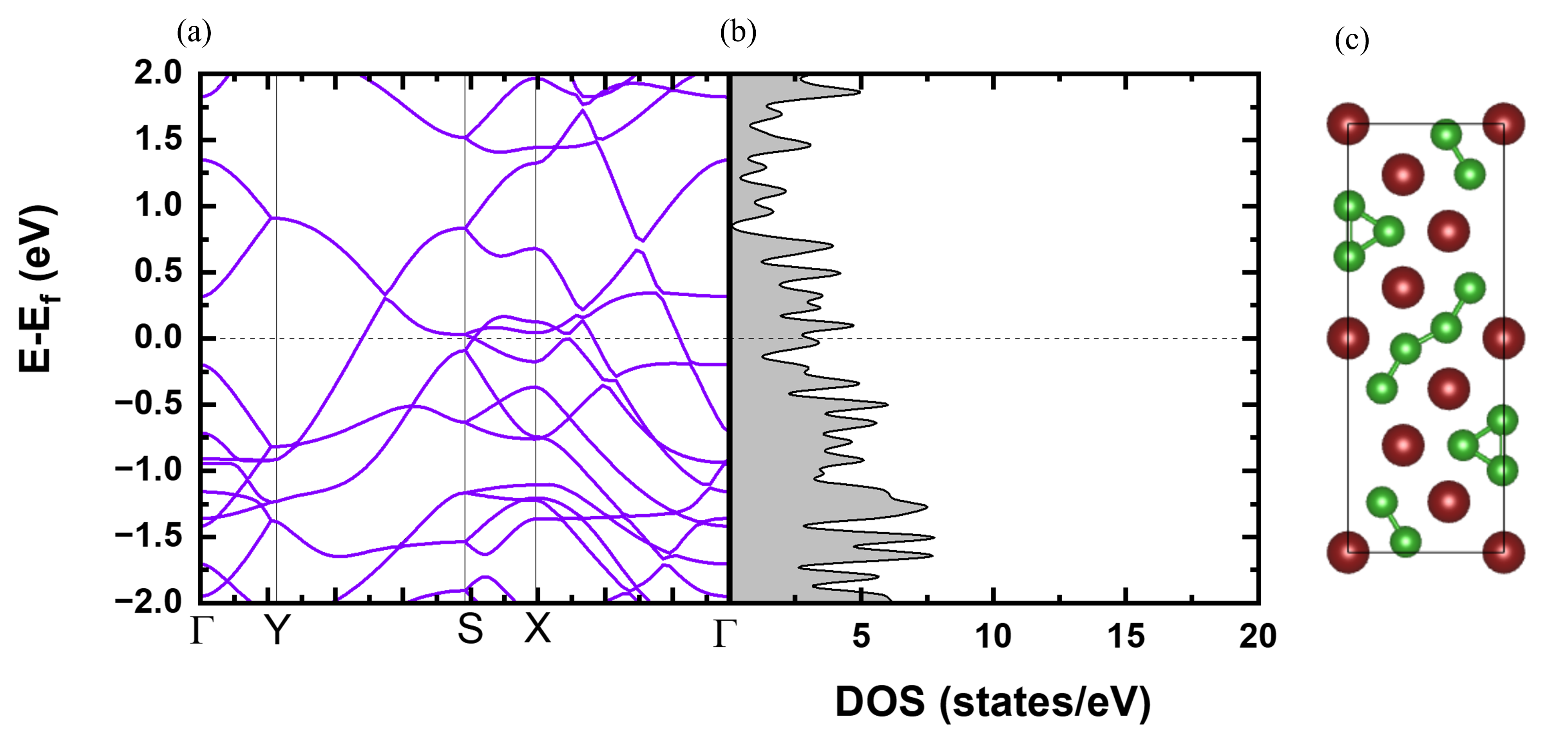}    
    \caption{Electronic properties of the pristine Cu$_{8}$B$_{14}$ monolayer: (a) electronic band structure and (b) total density of states (DOS) (c) the structural model.} 
    \label{fig:band_dos_pristine_1}%
\end{figure}

To simulate the two-dimensional copper boride structure, we employed two distinct unit cell configurations, as illustrated in Figure 1: (b) $4 \times 1$ supercell of Cu$_{8}$B$_{14}$, comprising 32 copper and 56 boron atoms, was constructed and subjected to ionic relaxation. The lattice parameters were adopted from prior experimental investigations of the copper boride monolayer~\cite{i31,i32}: for the $4 \times 1$ supercell, the parameters were $a = 21.8$~\AA, $b = 16.0$~\AA, and $\gamma = 70^\circ$, while the smaller rectangular unit cell (containing 8 Cu and 14 B atoms) used $a = 5.46$~\AA\ and $b = 15$~\AA. The calculated mean bond lengths were 2.14~\AA\ for Cu--B and 1.65~\AA\ for B--B, which align well with previously reported ranges of 2.04--2.35~\AA\ and 1.6--1.76~\AA, respectively~\cite{s3}.

The thermal stability of the system was evaluated by performing \textit{ab initio} molecular dynamics simulations at 400~K.
The simulation protocol involved an initial heating phase from 0 to 400~K over 3~ps, followed by a 5~ps equilibration period at 400~K in the canonical (NVT) ensemble with a time step of 1~fs. A Nosé--Hoover thermostat was utilised to enforce isothermal conditions~\cite{s4}. Figure~2 presents the fluctuations in total energy and bond lengths (Cu--Cu, B--B, Cu--B), along with a final structure. The slight variation in energy around the average energy suggests that the monolayer maintains its structural integrity at elevated temperatures.
The dynamical stability of the Cu$_8$B$_{14}$ monolayer was assessed using phonon dispersion calculations based on the finite-displacement method. As shown in Figure~3(a), no imaginary phonon frequencies are observed along the $\Gamma$--Y--S--X--$\Gamma$ high-symmetry path, supporting the dynamical stability of the relaxed structure within the harmonic approximation.
Additionally, we assessed the mechanical stability of the monolayer using the energy-strain method. The resulting elastic constants ($C_{ij}$), provided in Table~\ref{tab:elastic_constants} , which fulfills the Born Huang stability criteria ($C_{11}C_{22} > C_{12}^2$ and $C_{ij} > 0$), thereby confirming the mechanical stability of the Cu$_{8}$B$_{14}$ monolayer.

\begin{table}[h]
\centering
\caption{The elastic constants as calculated using the energy-strain method.}
\begin{tabular}{cc}
\hline
Elastic constants ($C_{ij}$) &  Magnitude(N/m) \\
\hline
$C_{11}$ &  130.11 \\
$C_{12}$ & 56.98 \\
$C_{22}$ & 129.74 \\
$C_{66}$ & 61.69 \\
\hline
\end{tabular}
\label{tab:elastic_constants}
\end{table}

To investigate the electronic structure of the material, we calculate the density of states (DOS) and band structure, which are depicted in Figure 4. As shown in Figure 4(b), the presence of finite states around the Fermi level indicates a metallic character, suggesting its excellent electrical conductivity. Furthermore, the electronic band dispersion displayed in Figure 4(a) is consistent with the DOS features and aligns well with previously reported data~\cite{s5}.

\subsection{Li adsorption on Cu$_{8}$B$_{14}$}

\begin{figure}
	\centering 
	\includegraphics[width=0.4\textwidth, angle=0]{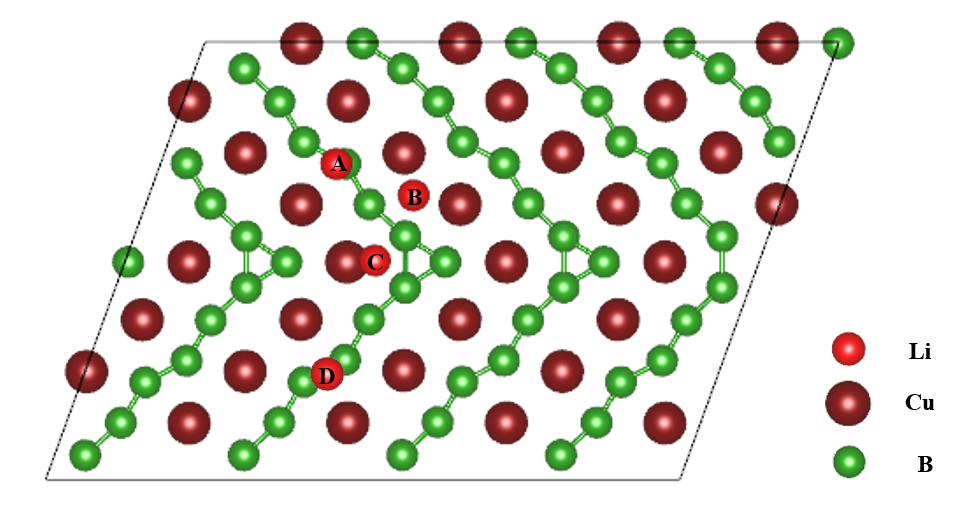}	
	\caption{Different adsorption sites (marked in red) for Li adsorption over the
Cu$_{8}$B$_{14}$ monolayer} 
	\label{fig:ads_sites}%
\end{figure}

To evaluate the lithium storage capability, we first investigated the adsorption properties of a single Li atom on the  Cu$_{8}$B$_{14}$ monolayer at different sites. Based on structural symmetry, eight distinct initial adsorption sites were identified. However, upon geometric relaxation, the Li atom converged to one of four stable configurations, which are depicted in Figure 5. Consequently, our analysis focused on these four unique sites. To determine the adsorption energy ($E_{\text{ads}}$), we utilized the following formulation:

\begin{equation}
    E_{\text{ads}} = \frac{E_{\text{nLi-Cu}_{8}\text{B}_{14}} - E_{\text{Cu}_{8}\text{B}_{14}} - nE_{\text{Li}}}{n}
    \label{eq:adsorption_energy}
\end{equation}

\noindent where $E_{\text{nLi-Cu}_{8}\text{B}_{14}}$ and $E_{\text{Cu}_{8}\text{B}_{14}}$ represent the total energies of the lithiated with n Li atoms and pristine monolayers, respectively. $E_{\text{Li}}$ stands for the energy per atom of lithium in the bulk state. The calculated adsorption energies at different sites, summarised in Table~3, indicate that Site B is the most energetically favourable, with an adsorption energy of $-1.1$~eV.

To further elucidate the electronic interactions at the most favourable adsorption site, we analysed the charge-transfer mechanism for a single Li atom adsorbed at this site. A quantitative assessment was performed using Bader charge analysis~\cite{a1}, which reveals a transfer of $0.89~e$ from the lithium atom to the Cu$_{8}$B$_{14}$ substrate. This substantial charge donation indicates strong ionic bonding between the adsorbate and the monolayer.

To visualize this redistribution, the charge density difference ($\Delta \rho$) was calculated according to the following equation and plotted in Figure 6 using VESTA~\cite{a2}:

\begin{equation}
    \Delta \rho = \rho_{\mathrm{Li-Cu}_{8}\mathrm{B}_{14}} - \left( \rho_{\mathrm{Cu}_{8}\mathrm{B}_{14}} + \rho_{\mathrm{Li}} \right)
    \label{eq:charge_diff}
\end{equation}

\noindent where $\rho_{\mathrm{Li-Cu}_{8}\mathrm{B}_{14}}$, $\rho_{\mathrm{Cu}_{8}\mathrm{B}_{14}}$, and $\rho_{\mathrm{Li}}$ represent the total charge densities of the lithiated system, the pristine Cu$_{8}$B$_{14}$ monolayer, and an isolated Li atom, respectively.

In Figure 6, yellow isosurfaces denote electron accumulation, while cyan regions indicate depletion. The plot clearly shows charge depletion near the Li atom and accumulation primarily on the boron-rich regions of the monolayer. These findings corroborate the Bader analysis, confirming a robust charge transfer process beneficial for battery performance.

\begin{figure}
    \centering 
    \includegraphics[width=0.5\textwidth, angle=0]{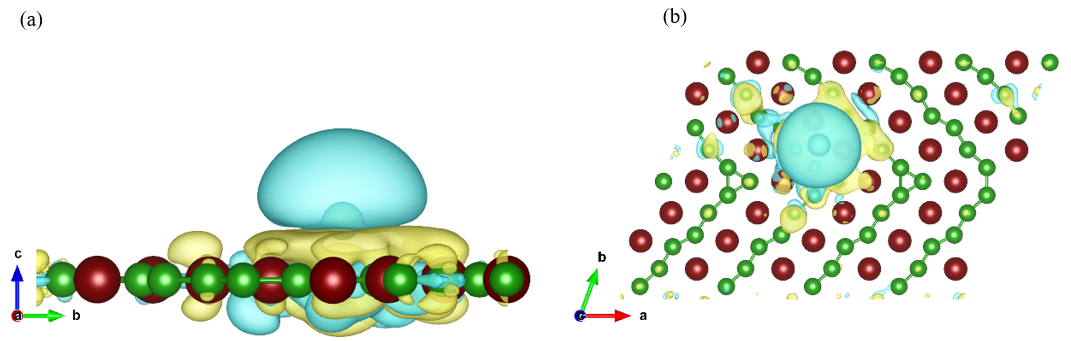}    
 \caption{(a) Side view and (b) top view of the charge density difference for a Li atom adsorbed at the most favorable site on the Cu$_8$B$_{14}$ monolayer (isosurface level = 0.0023 e/\AA$^3$).} 
    \label{fig:charge_difference}%
\end{figure}

Finally, to check the change in electronic properties with the insertion of Li atom, we have computed the band structure as well as the density of states calculation for a single Li adsorbed system
to understand the changes in its electronic properties.

\begin{figure}
	\centering 
	\includegraphics[width=0.55\textwidth, angle=0]{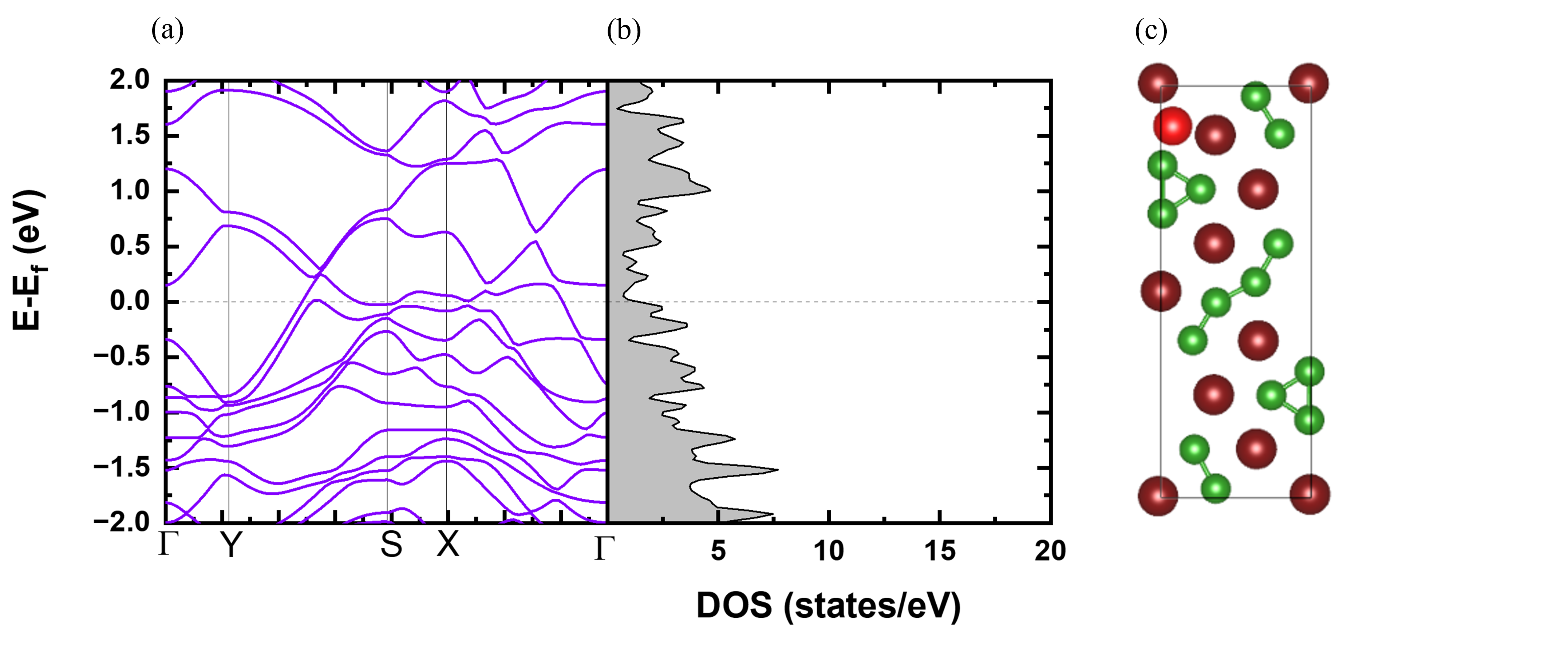}	
	\caption{Electronic properties of the Cu$_8$B$_{14}$ monolayer
with a single adsorbed Li atom: (a) electronic band structure,
(b) total density of states, and (c) relaxed adsorption geometry.
} 
	\label{fig:1li_band_dos}%
\end{figure}
As shown in Figure 7, both the density of states and the electronic band structure indicate that the system retains its conductive properties after Li adsorption.

\begin{table}[h]
\centering
\caption{Calculated adsorption energies of Li atoms at distinct symmetry sites.}
\begin{tabular}{lc}
\hline
Sites & Adsorption energy (eV) \\
\hline
 A & $-1.05$ \\
 B & $-1.10$ \\
 C & $-0.98$ \\
 D & $-1.05$ \\
\hline
\end{tabular}
\end{table}

\begin{table}[h]
\centering
\caption{Average Adsorption energy of lithium ion on Cu$_8$B$_{14}$ calculated using equation (3) with different concentration}
\begin{tabular}{lc}
\hline
\textbf{System} & $\mathbf{E_{ad}}$ \textbf{(eV)} \\
\hline
Cu$_8$B$_{14}$ + 4Li  & $-1.001$ \\
Cu$_8$B$_{14}$ + 8Li  & $-0.928$ \\
Cu$_8$B$_{14}$ + 12Li & $-0.800$ \\
Cu$_8$B$_{14}$ + 16Li & $-0.726$ \\
Cu$_8$B$_{14}$ + 20Li & $-0.677$ \\
Cu$_8$B$_{14}$ + 24Li & $-0.635$ \\
Cu$_8$B$_{14}$ + 28Li & $-0.603$ \\
Cu$_8$B$_{14}$ + 32Li & $-0.589$ \\
Cu$_8$B$_{14}$ + 36Li & $-0.570$ \\
Cu$_8$B$_{14}$ + 40Li & $-0.539$ \\
Cu$_8$B$_{14}$ + 42Li & $-0.532$ \\
\hline
\end{tabular}
\label{tab:adsorption_energy}
\end{table}

\subsection{Voltage Profile and Storage Capacity}

\begin{figure}
	\centering 
	\includegraphics[width=0.55\textwidth, angle=0]{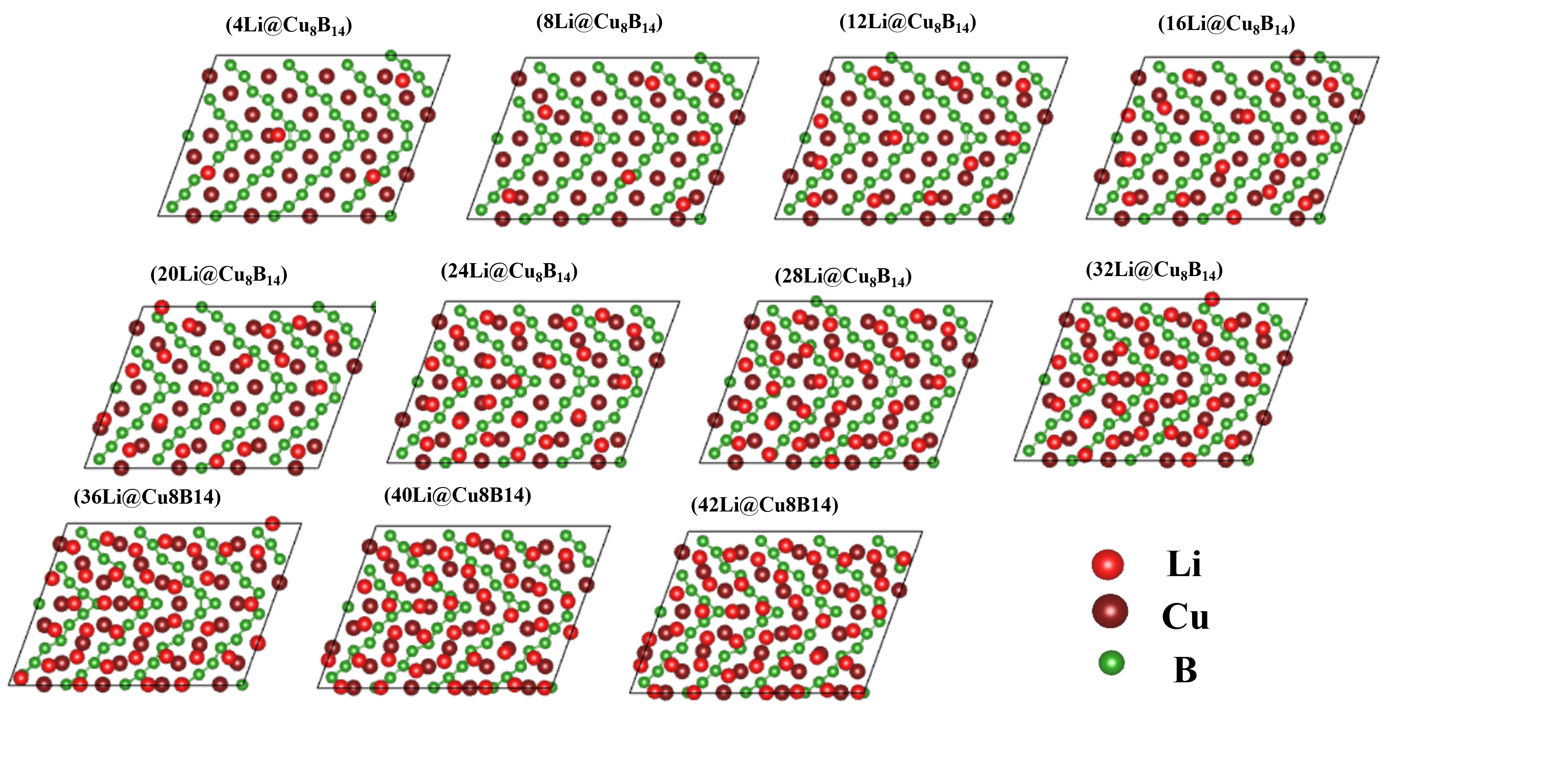}	
	\caption{The incremental adsorption of Li atoms on the Cu$_{8}$B$_{14}$ monolayer} 
	\label{fig:specific_capacity}%
\end{figure}

To compute the maximum theoretical specific capacity, lithium atoms were incrementally introduced to the Cu$_{8}$B$_{14}$ monolayer. The loading process was governed by four limiting criteria: (a) the formation of Li clusters (interatomic distances $< 2.67$~\A)\cite{d2}, (b) the desorption of Li atoms from the surface, (c) a transition to positive or negligible adsorption energies, and (d) significant structural distortion of the host monolayer.

In practice, Li atoms were added in steps of four at favourable adsorption sites. The loading was halted when the Li--Li distances dropped below 2.67~\AA, ensuring no metallic clustering occurred. Under these conditions, the monolayer successfully accommodated 42 Li atoms on a single side, corresponding to a specific capacity of 427~mAh~g$^{-1}$. If an interlayer spacing of 9~\AA\ is maintained, adsorption on both surfaces could double the capacity to 854~mAh~g$^{-1}$. The theoretical specific capacity ($C$) was calculated using:

\begin{equation}
    C = \frac{n F}{M_{\text{Cu}_{8}\text{B}_{14}}}
    \label{eq:capacity}
\end{equation}

\noindent where $n$ represents the maximum number of Li atoms adsorbed, $F$ is the Faraday constant (26,801~mAh~mol$^{-1}$), and $M_{\text{Cu}_{8}\text{B}_{14}}$ is the molar mass of the supercell. Throughout the lithiation process, the average adsorption energy remained within the favorable range of $-1.0$~eV to $-0.53$~eV, confirming thermodynamically stable storage. Figure 8 illustrates the optimized structures at various loading stages, while the corresponding average adsorption energies are given in Table~4.

The storage capacity of Cu$_{8}$B$_{14}$ surpasses that of pristine graphene and many other reported materials, as summarised in Table~5.

Along with specific capacity, another property for battery electrode is the open-circuit voltage (OCV), which is a pivotal determinant of overall battery efficacy, directly influencing cyclic efficiency, operational safety and long-term stability. A consistent OCV profile suggests that the anode material preserves its structural integrity during lithiation and delithiation, thereby mitigating severe structural breakdown or adverse phase transitions. The anode potential should ideally be low and positive, in the range of 0.1 to 1.0 V ~\cite{v1}. Potentials below this window risk lithium plating on the electrode surface, which depletes the active lithium count and fosters dendrite growth, compromising cell safety~\cite{a3}. In this work, we determined the OCV profile of the Cu$_{8}$B$_{14}$ monolayer utilising convex-hull analysis~\cite{a4}. The relative formation energy at each Li loading was calculated with respect
to the pristine and fully lithiated end members. The normalized lithiation
fraction was defined as $\xi=n/n_{\mathrm{max}}$, where $n$ is the number
of adsorbed Li atoms and $n_{\mathrm{max}}$ is the maximum Li loading.
The formation energy at a given lithiation fraction $\xi$ was then calculated
as

\begin{equation}
E_{\mathrm{form}}(\xi)
=
E_{n}
-
(1-\xi)E_{0}
-
\xi E_{n_{\mathrm{max}}},
\label{eq:formation_energy}
\end{equation}

\noindent where $E_{n}$ is the total energy of the Cu$_8$B$_{14}$ monolayer
containing $n$ adsorbed Li atoms, $E_{0}$ is the total energy of the pristine
Cu$_8$B$_{14}$ monolayer, and $E_{n_{\mathrm{max}}}$ is the total energy of
the fully lithiated configuration containing $n_{\mathrm{max}}$ Li atoms.
With this definition, the pristine ($\xi=0$) and fully lithiated ($\xi=1$)
end members have zero formation energy, while intermediate configurations
lying on the lower convex hull correspond to the thermodynamically preferred
lithiation states relative to the two end members.

\begin{figure}
    \centering
    \includegraphics[width=0.55\textwidth, angle=0]{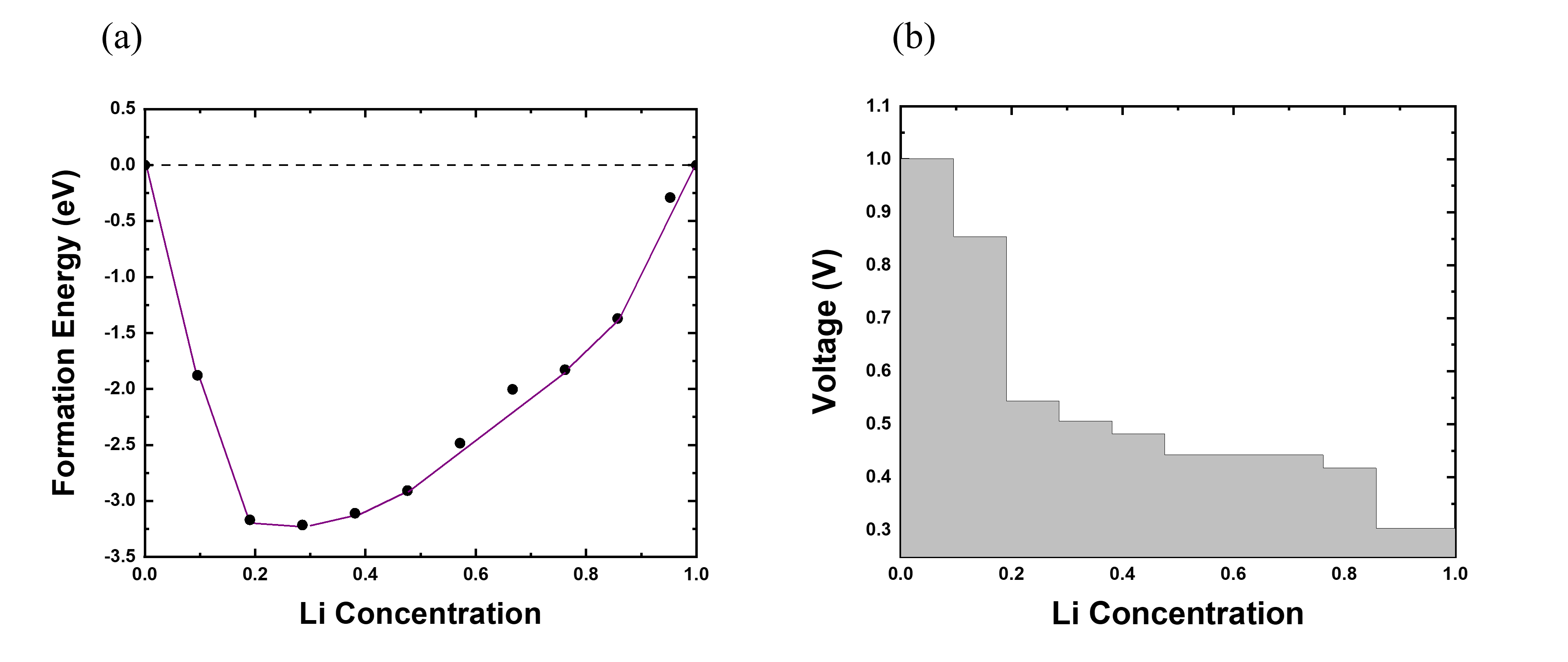}
    \caption{(a) Convex hull as a function of the normalized Li concentration
    and (b) the corresponding voltage profile of Li-adsorbed
    Cu$_8$B$_{14}$.}
    \label{fig:convex_hull_voltage}
\end{figure}

A lower convex hull was constructed from the calculated formation energies
to identify the thermodynamically preferred intermediate lithiation
configurations, as shown in Figure~\ref{fig:convex_hull_voltage}(a).
The average open-circuit voltage between two adjacent stable configurations
on the convex hull was estimated from the corresponding DFT total-energy
difference according to

\begin{equation}
V(n_a,n_b)
=
-\frac{
E_{n_a}-E_{n_b}-(n_a-n_b)E_{\mathrm{Li}}^{\mathrm{bulk}}
}{
(n_a-n_b)e
},
\label{eq:voltage}
\end{equation}

\noindent where $n_a$ and $n_b$ are the numbers of adsorbed Li atoms in two
adjacent stable configurations, with $n_a>n_b$, and $E_{n_a}$ and $E_{n_b}$
are their corresponding total energies. $E_{\mathrm{Li}}^{\mathrm{bulk}}$
denotes the energy per atom of bulk Li, and $e$ is the elementary charge.
The $pV$ and entropic contributions were neglected, such that the Gibbs
free-energy change was approximated by the DFT total-energy difference.
The resulting voltage profile is shown in
Figure~\ref{fig:convex_hull_voltage}(b).

By averaging the voltage plateaus across the entire lithiation range, the mean open-circuit voltage for the Cu$_{8}$B$_{14}$ anode was determined to be 0.53~V. This value lies well within the optimal potential window for anode materials (0.1--1.0~V), confirming that the Cu$_{8}$B$_{14}$ monolayer is a promising candidate for safe and efficient lithium-ion battery applications.

\begin{table}[t]
\centering
\caption{Comparative data of storage capacity and Diffusion barrier for electrode materials.}
\label{tab:table3}
\renewcommand{\arraystretch}{1.1}

\resizebox{\columnwidth}{!}{%
\begin{tabular}{@{}lccc@{}}
\toprule
\textbf{Materials} &
\textbf{Lithium storage capability (mAh g$^{-1}$)} &
\textbf{Diffusion Barrier (eV)} &
\textbf{Reference} \\
\midrule
VB        & 390.168  &  0.264       & \cite{T1} \\
ScB          & 427.373  & 0.108    & \cite{T1} \\
H-Mo$_{2}$B          & 74.18  & 0.050        & \cite{T2} \\
Ti$_{2}$B                 & 503.1  & 0.017--0.0235 & \cite{T3} \\
Mo$_{2}$B$_{2}$            & 444 & 0.27    & \cite{T4} \\
Ti$_{2}$B$_{2}$     & 456  & 0.017      & \cite{T5} \\
TiB          & 408.4  & 0.105      & \cite{T1} \\
Ti$_{3}$C$_{2}$          &  320  & 0.28     & \cite{T7} \\
T-Mo$_{2}$B             & 264  & 0.037      & \cite{T2} \\
Cu$_{8}$B$_{14}$      & 427 & 0.32--0.41      & Present Work \\
\bottomrule
\end{tabular}%
}
\end{table}

\subsection{Li migration over the Cu$_{8}$B$_{14}$ monolayer}

\begin{figure}
	\centering 
	\includegraphics[width=0.5\textwidth, angle=0]{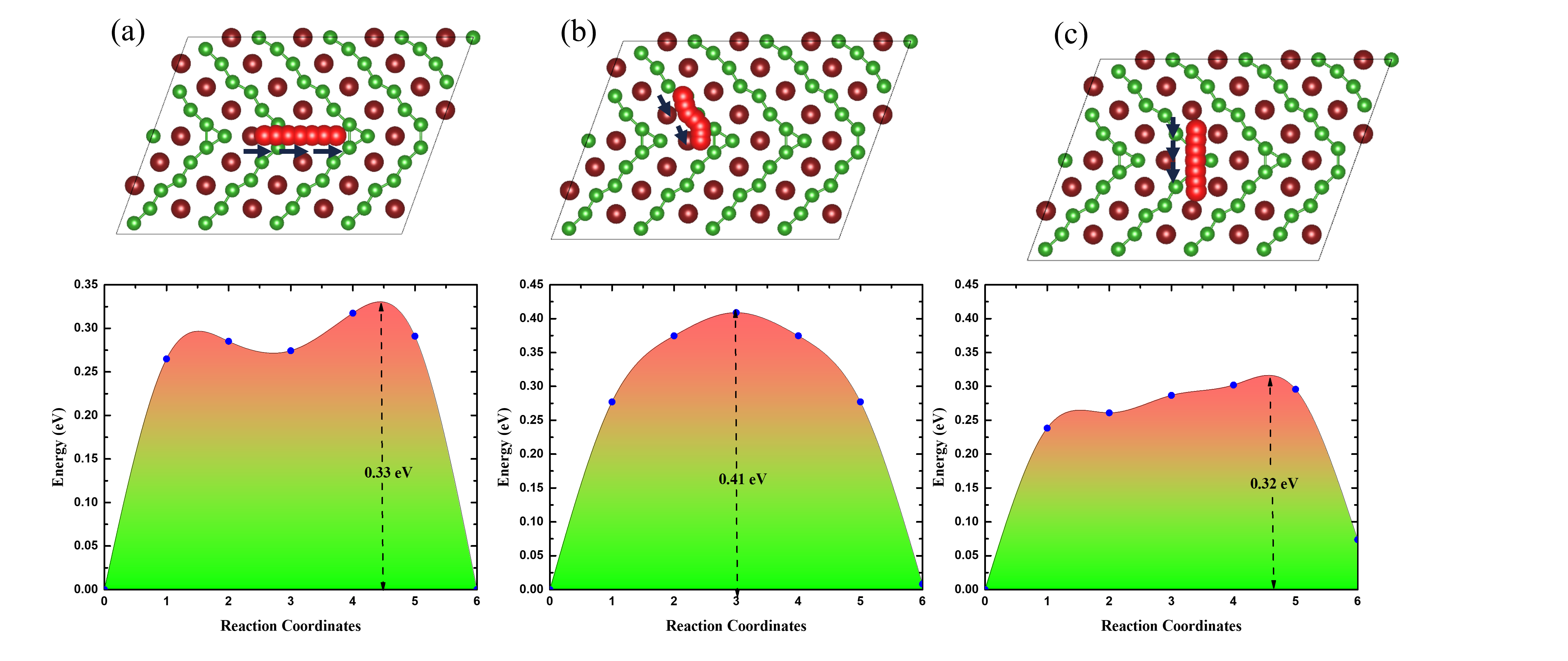}	
	\caption{The migration pathway and diffusion energy barrier for Li atom along (a) Path 1, (b) Path 2, and (c) Path 3 over the monolayer
} 
	\label{fig:band_dos_pristine_1}%
\end{figure}

A critical factor determining anode performance is the ease of lithium ion migration through the electrode material during charge-discharge cycles, which is quantified by the diffusion energy barrier. To assess this, we employed the nudged elastic band (NEB) method~\cite{m8} to calculate the migration energy barrier across the Cu$_{8}$B$_{14}$ monolayer.

We first identified potential surface diffusion pathways and subsequently computed the migration energy barriers along three distinct routes, as illustrated in Figure 10. The first pathway (Figure 10 (a)) shows a migration barrier of 0.33~eV, while the second route (Figure 10 (b)) presents a barrier of 0.41~eV. The third pathway (Figure 10 c) displays the lowest barrier at 0.32~eV . These migration barriers align perfectly with the established 0.33~eV for a pristine graphene monolayer~\cite{d1}, indicating that the Cu$_{8}$B$_{14}$ monolayer possesses favourable diffusion characteristics. Consequently, the Cu$_{8}$B$_{14}$ monolayer offers fast lithium kinetics, making it highly viable for high-power applications and a promising candidate for further investigation.

\subsection{AIMD analysis of the fully lithiated monolayer}

\begin{figure}
	\centering 
	\includegraphics[width=0.52\textwidth, angle=0]{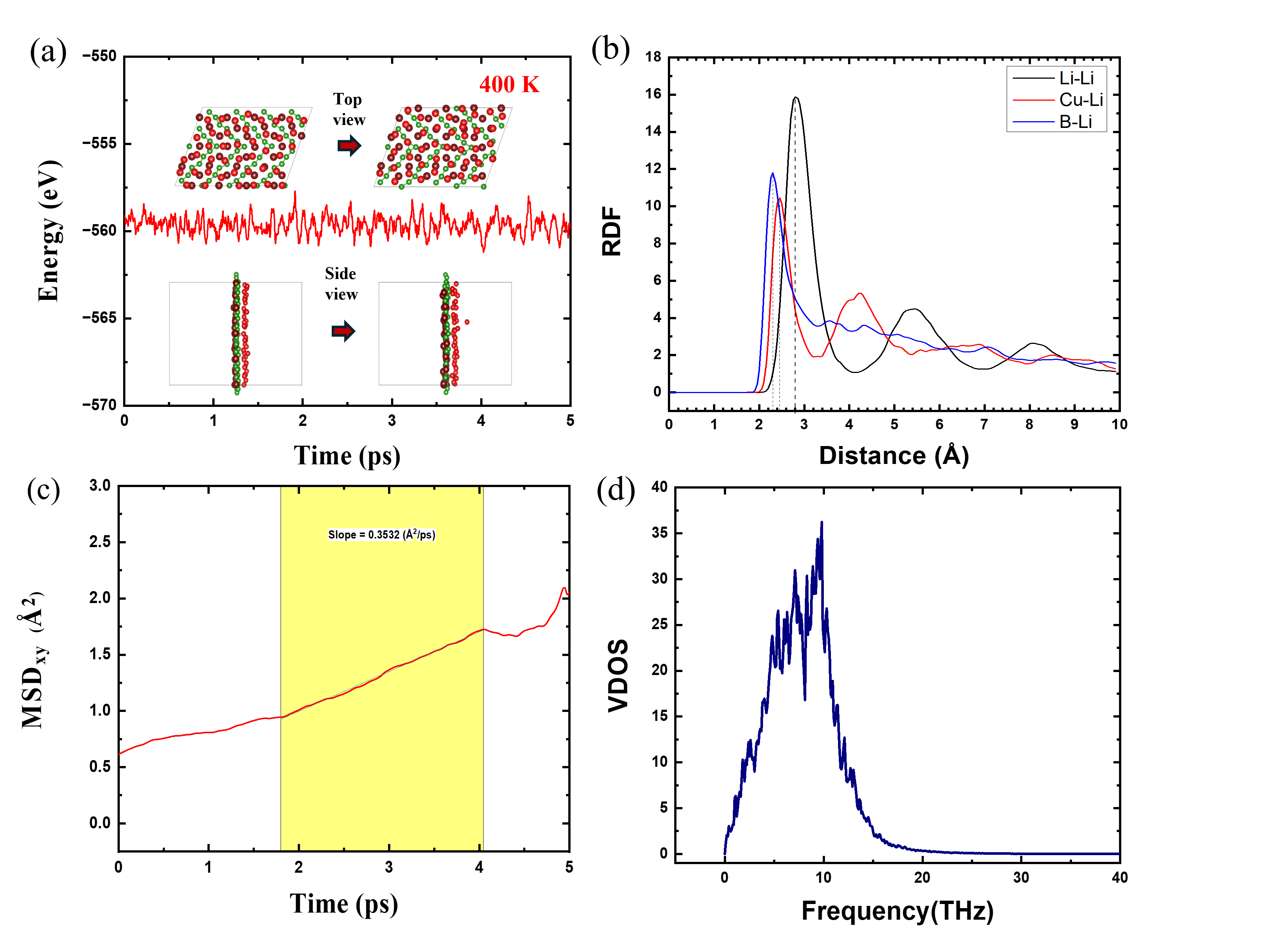}	
	\caption{AIMD simulation on fully lithiated Cu$_{8}$B$_{14}$ (a) Total energy variation with Time, (b) Radial distribution function for Li-Li, Cu-Li and B-Li, and (c) 2d MSD for Li atoms along with the linear region.
(d) Vibrational density of states of Li} 
	\label{fig:msd_aimd}%
\end{figure}
To assess the thermal stability and Li dynamics of the fully
lithiated Cu$_8$B$_{14}$ monolayer, AIMD simulations were performed
at $400~\mathrm{K}$. As shown in Figure~11(a), the total energy
remains bounded with relatively small fluctuations throughout the
trajectory, while the Cu--B framework preserves its structural
integrity. These results confirm the thermal stability of the fully
lithiated monolayer under the simulated conditions.

The spatial distribution of Li atoms was further examined using
radial distribution functions (RDFs)~\cite{m12}. Figure~11(b)
presents the Li--Li, B--Li, and Cu--Li pair correlations. The first
Li--Li coordination peak occurs at approximately $2.8~\mathrm{\AA}$,
which is larger than the Li--Li separation of
$2.67~\mathrm{\AA}$ in a Li$_2$ molecule~\cite{d2}. The absence of
pronounced Li--Li correlations at shorter separations indicates that
compact Li clustering is suppressed during the AIMD trajectory.
In contrast, the B--Li and Cu--Li RDFs exhibit prominent short-range
features, showing that Li remains preferentially coordinated with the
Cu--B host framework rather than aggregating with neighbouring Li
atoms.

To assess Li mobility, the in-plane mean-squared displacement (MSD)
was extracted from the AIMD trajectory, as shown in
Figure~11(c). An approximately linear increase in the MSD is observed
over part of the trajectory, from which an in-plane Li tracer
diffusion coefficient of $8.83\times10^{-6}~\mathrm{cm^2\,s^{-1}}$
at $400~\mathrm{K}$ was estimated using the Einstein relation.
This value exceeds the upper end of the range typically reported
for electrode materials
($10^{-10}$--$10^{-6}~\mathrm{cm^2\,s^{-1}}$)~\cite{d3},
indicating rapid Li mobility in the fully lithiated
Cu$_8$B$_{14}$ monolayer. Nevertheless, because the AIMD trajectory
extends over only a few picoseconds, the resulting value should be
regarded as a short-time estimate rather than a converged long-time
diffusion coefficient. To obtain a statistically more robust
description of Li transport over extended time scales,
nanosecond-scale MLIP-MD simulations were therefore performed, as
discussed in Section 3.7.

Furthermore, the Li vibrational density of states (VDOS) was obtained
from the Fourier transform of the Li velocity autocorrelation
function, as shown in Figure~11(d). The spectrum exhibits a broad
distribution of vibrational frequencies, with pronounced spectral
weight centred around approximately $9.8~\mathrm{THz}$. Such spectral
broadening is consistent with Li atoms sampling a range of local
coordination environments and undergoing anharmonic motion within the
Cu$_8$B$_{14}$ framework. The broad vibrational response therefore
complements the AIMD-based MSD analysis by demonstrating substantial
dynamical activity of Li at $400~\mathrm{K}$, although the VDOS itself
does not provide a direct measure of long-range diffusion.

\subsection{Line Defect Analysis}
Structural and compositional defects are ubiquitous in synthesized 2D materials and can drastically alter their electrochemical behavior. To accurately evaluate the practical anode performance of copper boride, we introduced a line defect into the pristine structure Figure 12. We based our model on the specific line-defect configuration recently identified experimentally by Li et al. (2025)~\cite{i32}. Assessing the influence of this defect is crucial, as extended structural anomalies inevitably dictate the macroscopic capacity, ion diffusion kinetics, and overall stability of the material. Here in this defect analysis, we studied three important parameters: thermodynamic stability, specific capacity and diffusion kinetics.

The Li-storage capability of the line-defected monolayer was evaluated
using the same lithiation criterion employed for the pristine system. The defect monolayer accommodated a maximum of 27 Li atoms on a single surface, yielding a theoretical specific capacity of approximately 400 mAh/g. Although slightly lower than the
$427~\mathrm{mAh\,g^{-1}}$ obtained for pristine Cu$_8$B$_{14}$,
the capacity remains comparable to that of commercial graphite
($372~\mathrm{mAh\,g^{-1}}$). The reduction could potentially originates from
the modified Cu/B stoichiometry of the defect structure
(Cu$_{11}$B$_{19}$), which increases the relative concentration of
the heavier Cu to the host mass.

The thermal stability of the fully lithiated line-defected
Cu--B monolayer was examined using AIMD at $400~\mathrm{K}$.
As shown in Figure~13(a), the total energy remains bounded with
relatively small fluctuations throughout the trajectory, with a
standard deviation of approximately $0.42~\mathrm{eV}$. The Cu--B
framework also preserves its structural integrity without noticeable
bond breaking or lattice reconstruction, demonstrating the thermal
robustness of the lithiated defect structure under the simulated
conditions.

The in-plane Li mean-squared displacement is shown in
Figure~13(b). Following an initial non-diffusive regime, the MSD
exhibits an approximately increasing region during the latter part
of the AIMD trajectory. A linear fit to this interval gives an
estimated in-plane Li tracer diffusion coefficient of approximately
$9.04\times10^{-5}~\mathrm{cm^2\,s^{-1}}$ at $400~\mathrm{K}$.
This short-time estimate is higher than that obtained for the
pristine lithiated monolayer, suggesting enhanced Li mobility in
the presence of the line defect. However, owing to the limited
picosecond time scale and the pronounced temporal variation of the
MSD, this value should not be interpreted as a converged long-time
diffusion coefficient.

To evaluate the specific influence of these structural imperfections on ionic mobility, we calculated the lithium diffusion barrier along the designated migration trajectory using the nudged elastic band method (Figure 14). The calculated activation barrier is 0.21 eV, This reduction indicates that the line defect locally
facilitates Li migration and is consistent with the enhanced
short-time Li mobility inferred from the AIMD trajectory. However,
because the AIMD-derived diffusivity is obtained from a picosecond
trajectory, the extent to which this local reduction in the migration
barrier translates into sustained long-time transport is examined
more rigorously using the nanosecond-scale MLIP-MD simulations in the
following section.

\begin{figure}
	\centering
	\includegraphics[width=0.5\textwidth, angle=0]{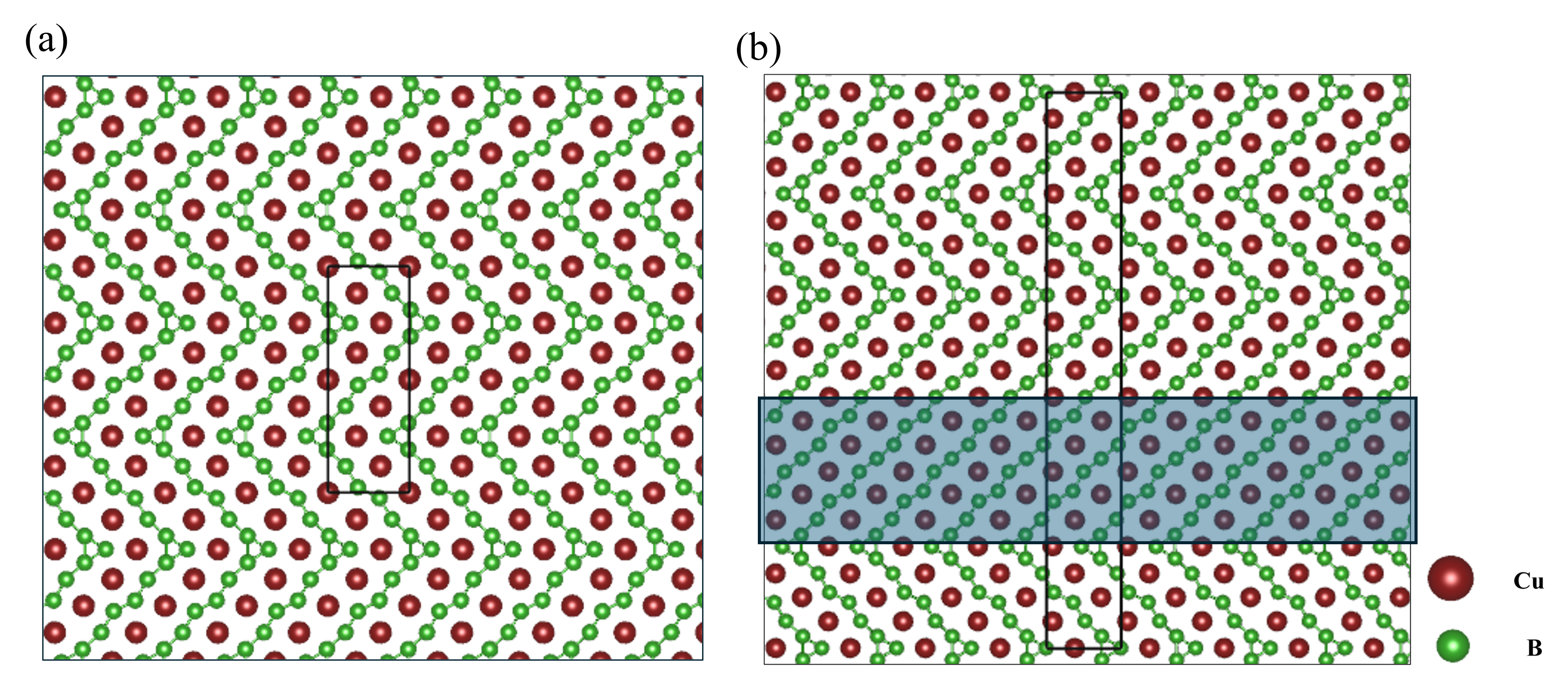}	
	\caption{Top view of the (a) pristine and (b) line-defected copper boride monolayers. Solid black rectangles indicate the primitive unit cells of both monolayers. The shaded blue region highlights the line defect.}
	\label{fig:cub_structure_defect}
\end{figure}

\begin{figure}
	\centering 
	\includegraphics[width=0.55\textwidth, angle=0]{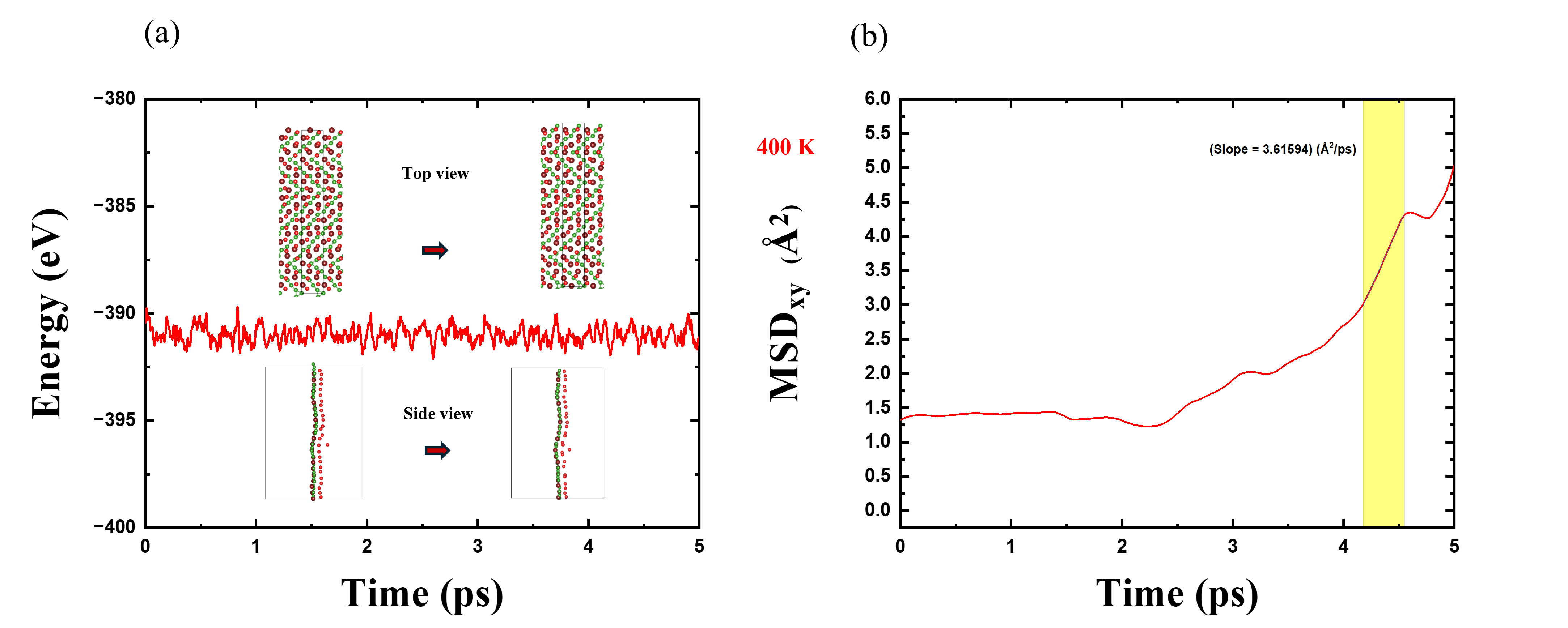}	
	\caption{(a) Total energy variation with time for fully lithiated defect structure (b) corresponding 2d MSD vs lag-time plot for Li atoms} 
	\label{fig:msd_aimd}%
\end{figure}

\begin{figure}
	\centering
	\includegraphics[width=0.48\textwidth, angle=0]{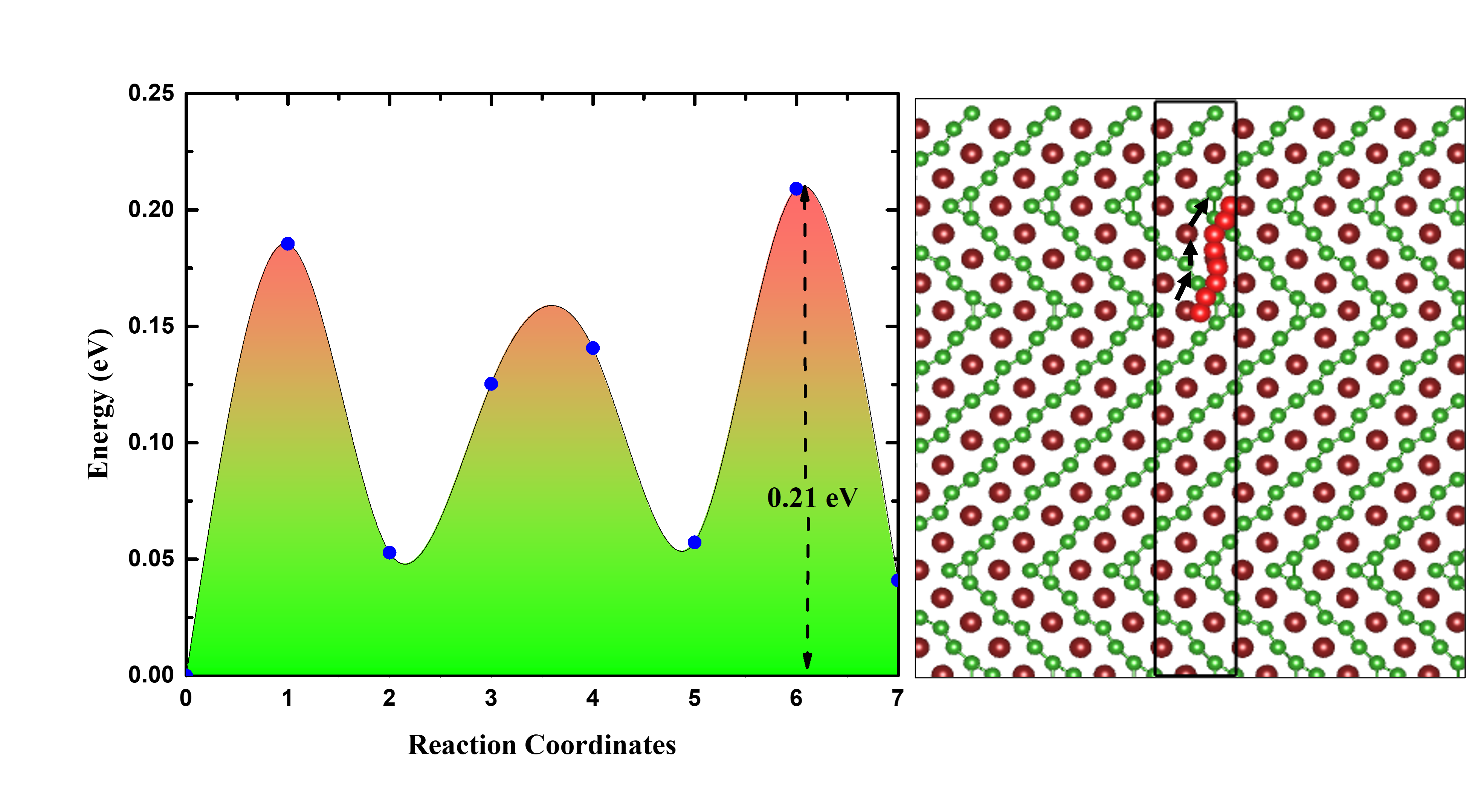}	
	\caption{The migration pathway and diffusion energy barrier for line defect structure} 
	\label{fig:msd_aimd}%
\end{figure}

\subsection{Nanosecond-scale Li transport from MLIP-MD}
\label{sec:mlip_transport}

\begin{figure*}
	\centering
	\includegraphics[width=0.95\textwidth, angle=0]{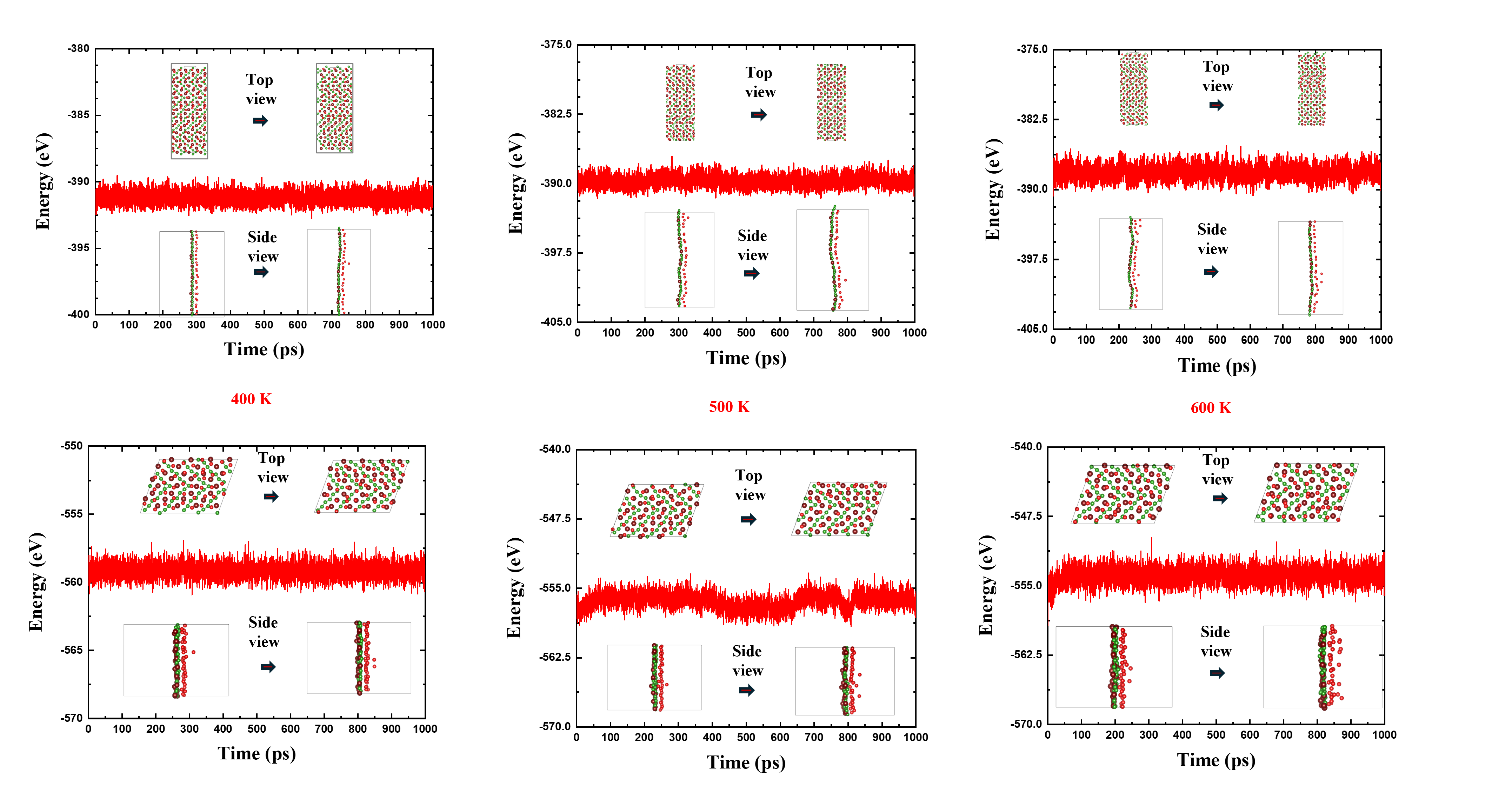}	
	\caption{Energy evolution during nanosecond-scale
MLIP-MD simulations of the fully lithiated pristine and
line-defected Cu--B monolayers at 400, 500, and 600~K.}
	\label{fig:ml-ip-all}%
\end{figure*}

\begin{figure}
	\centering
	\includegraphics[width=0.45\textwidth, angle=0]{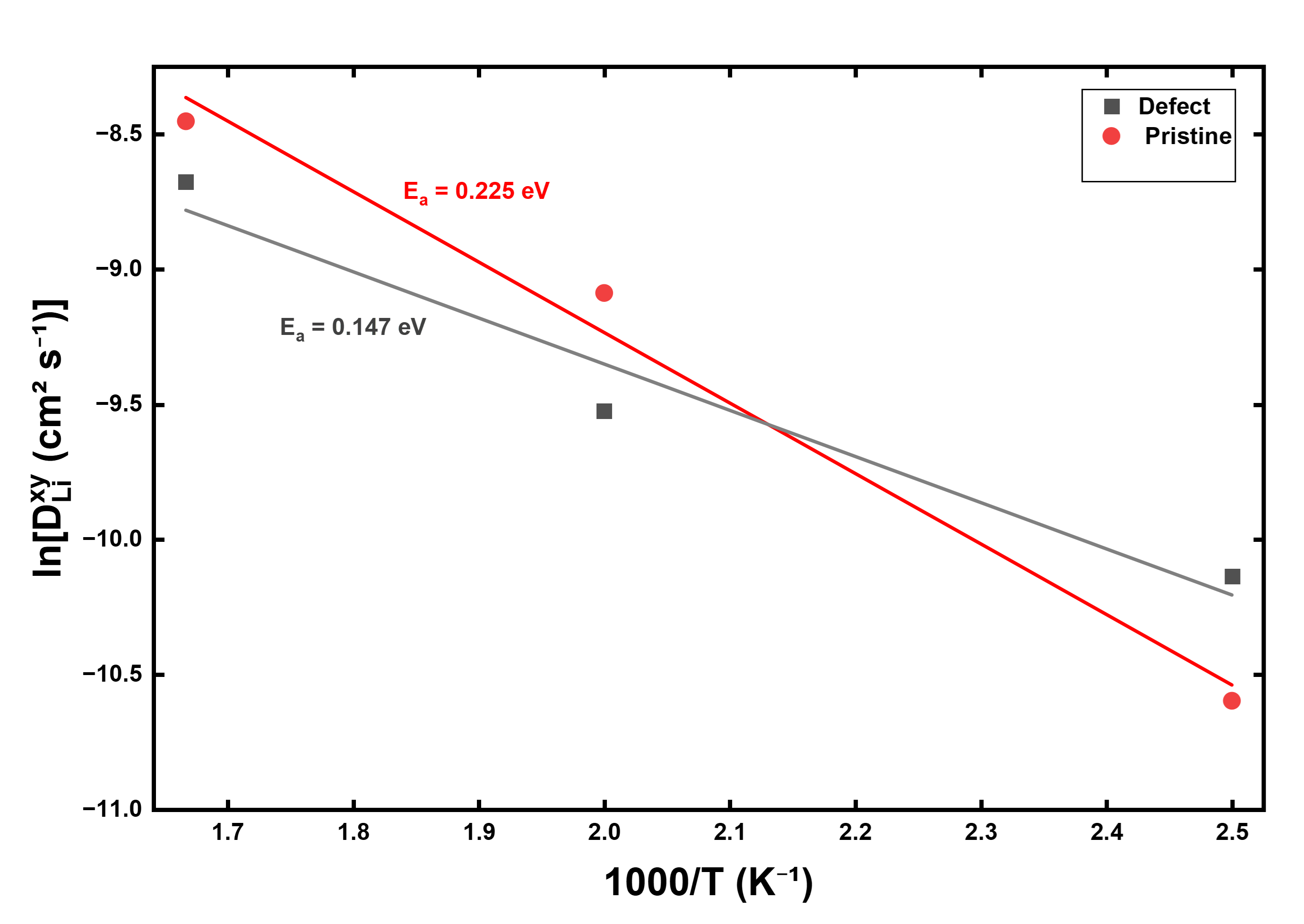}	
	\caption{ Arrhenius behaviour for line defect and pristine structures at temperatures [400 K ,500 K, 600 K]} 
	\label{fig:arrhenius}%
\end{figure}

The short time scale accessible to AIMD provides only a limited
description of Li transport, particularly because individual
migration events can strongly influence diffusivities extracted
from picosecond trajectories. We therefore employed the
fine-tuned MACE-MP-0 potential to extend the simulations to the
nanosecond regime and to examine Li transport in the fully
lithiated pristine and line-defected Cu--B monolayers over a
broader temperature range. Before analysing the long-time
trajectories, the reliability of the potential was assessed against
DFT reference data. As summarized in Table~\ref{tab:mace_validation},
On the held-out test set, system-specific fine-tuning reduced
the energy RMSE from 187.78 to 11.18~meV per atom.
The force RMSE decreased from 273.03 to
58.22~meV\,\AA$^{-1}$. In addition, evaluation against an independent
defect+27Li AIMD trajectory, which was excluded from model
development, yielded a Li-force RMSE of
$46.22~\mathrm{meV\,\AA^{-1}}$ and a mean Li-force cosine
similarity of $0.968$. The agreement obtained for configurations
outside the training, validation, and test sets provides support
for using the fine-tuned potential to investigate long-time Li
dynamics.

Prior to analysing Li diffusion, the energetic and structural
stability of the fully lithiated systems was examined throughout
the MLIP-MD trajectories. Figure~\ref{fig:ml-ip-all} shows
the potential-energy evolution of the pristine and line-defected
monolayers at 400, 500, and 600~K. In each case, the potential
energy remains bounded throughout the nanosecond trajectory,
without a systematic drift indicative of progressive structural
instability. Inspection of the corresponding atomic configurations
further shows that the Cu--B framework remains intact over the
entire simulation interval, with no persistent lattice
fragmentation or large-scale reconstruction observed even at
600~K. The increase in the amplitude of the instantaneous energy
fluctuations with temperature is consistent with the larger
thermal motion sampled at elevated temperature rather than a
loss of framework stability. These results indicate that both the
pristine and line-defected lithiated monolayers remain
structurally stable over the temperature range used for the
subsequent diffusion analysis.

The time-origin-averaged in-plane Li MSD curves obtained from the
nanosecond trajectories are provided in Fig.~S1. For both systems,
the MSD increases systematically with temperature, demonstrating the
enhanced long-range Li mobility at elevated temperature. The resulting
in-plane tracer diffusion coefficients for the line-defected structure
are $3.97\times10^{-5}$, $7.30\times10^{-5}$, and
$1.71\times10^{-4}~\mathrm{cm^{2}\,s^{-1}}$ at 400, 500, and
600~K, respectively. The corresponding values for the pristine
monolayer are $2.50\times10^{-5}$, $1.13\times10^{-4}$, and
$2.14\times10^{-4}~\mathrm{cm^{2}\,s^{-1}}$, respectively.
These nanosecond-scale results reveal that the influence of the line
defect on Li transport is temperature dependent. At 400~K, the
diffusivity of the line-defected system is approximately 1.6 times
that of the pristine monolayer, whereas at 500 and 600~K the pristine
system exhibits the larger tracer diffusivity. Thus, although the
line defect can facilitate Li motion under particular conditions, it
does not produce a uniform enhancement of the ensemble-averaged
long-time diffusivity across the investigated temperature range.

The temperature dependence of the Li tracer diffusivity was further
examined using the Arrhenius relation in
Eq.~\ref{eq:arrhenius_diffusion}. As shown in
Fig.~\ref{fig:arrhenius}, linear fits of
$\ln D_{\mathrm{Li}}^{xy}$ against $1000/T$ yield effective
activation energies of $0.147~\mathrm{eV}$ for the line-defected
system and $0.225~\mathrm{eV}$ for the pristine monolayer, with
coefficients of determination of $R^{2}=0.957$ and $0.987$,
respectively. The corresponding pre-exponential factors are
$2.65\times10^{-3}$ and
$1.80\times10^{-2}~\mathrm{cm^{2}\,s^{-1}}$ for the defect and
pristine systems, respectively. The lower effective activation
energy obtained for the line-defected structure indicates a weaker
temperature dependence of its long-time Li transport over the
400--600~K range. However, the substantially different
pre-exponential factors result in crossing transport trends, such
that the lower activation energy of the defect system does not imply
a larger diffusivity at every temperature. Accordingly, the
Arrhenius analysis is more appropriately interpreted as evidence
that the line defect modifies the effective transport landscape,
rather than providing a uniform reduction in the resistance to
long-range Li diffusion.

\section{Conclusions}

In summary, we investigated the structural, electronic, and
electrochemical properties of the experimentally realized
Cu$_8$B$_{14}$ monolayer as a potential anode material for
Li-ion batteries using first-principles calculations combined with
machine-learning-interatomic-potential molecular dynamics. The
pristine monolayer remains metallic upon lithiation and exhibits
favourable Li adsorption, a single-surface theoretical capacity of
$427~\mathrm{mAh\,g^{-1}}$, and an average open-circuit voltage of
approximately $0.53~\mathrm{V}$. The lowest Li migration barrier
identified for the pristine surface is $0.32~\mathrm{eV}$. AIMD
simulations further indicate that the fully lithiated structure
retains its integrity at 400~K, although diffusivities extracted from
the picosecond trajectories are necessarily short-time estimates.

To access long-time Li transport, a MACE-MP-0 foundation model was
fine-tuned against system-specific DFT energies and forces and
subsequently employed for nanosecond-scale molecular dynamics. The
resulting trajectories show that both pristine and line-defected
lithiated monolayers remain structurally stable over the investigated
400--600~K range. For the experimentally motivated line-defect
structure, the single-surface capacity remains approximately
$400~\mathrm{mAh\,g^{-1}}$, while the local NEB migration barrier is
reduced to $0.21~\mathrm{eV}$. The long-time transport analysis,
however, reveals a more nuanced influence of the defect. At 400~K,
the line-defected system exhibits a higher tracer diffusivity than the
pristine monolayer, whereas the pristine system becomes more diffusive
at 500 and 600~K. Arrhenius analysis yields effective activation
energies of $0.147~\mathrm{eV}$ for the line-defected system and
$0.225~\mathrm{eV}$ for the pristine monolayer. These results show
that the line defect modifies the Li-transport landscape rather than
producing a uniform enhancement of long-range diffusion across all
temperatures. Overall, Cu$_8$B$_{14}$ combines metallic electronic
character, competitive Li-storage capacity, structural robustness,
and favourable Li mobility, supporting further experimental
investigation of pristine and defect-containing copper-boride
monolayers as Li-ion battery anodes.

\section*{Acknowledgements}

The authors would like to thank  the Indian Institute of Technology (Indian School of Mines), Dhanbad for providing the computational support and research facilities.

\bibliographystyle{elsarticle-num} 
\bibliography{References}

\end{document}